\documentclass[10pt,prd,aps,amssymb,amsmath,tightenlines,showpacs]{revtex4}

\usepackage{amsmath}
\usepackage{amssymb}

\usepackage{graphicx}
\usepackage[T1]{fontenc}
\usepackage[latin9]{inputenc}
\usepackage{color}

\usepackage{esint}

\usepackage{hyperref} 

\usepackage{selinput}
\SelectInputMappings{%
  adieresis={ä},
  eacute={é},
  Lcaron={Ľ},
}

\newcommand{\cC}{\ensuremath{\mathcal{C}}}

\newcommand{\cT}{\ensuremath{\mathcal{T}}}

\newcommand{\beq}{\begin{equation}}
\newcommand{\eeq}{\end{equation}}
\newcommand{\beqa}{\begin{eqnarray}}
\newcommand{\eeqa}{\end{eqnarray}}
\begin{document}

\title{
Non-autonomous Hamiltonian quantum systems, operator equations  \\
 and representations of Bender-Dunne Weyl ordered basis \\ under 
 time-dependent canonical transformations.
}

\author{Mariagiovanna Gianfreda$^a$}
\email{mariagiovanna.gianfreda@gmail.com}
\author{Giulio Landolfi$^{b\,c}$}\email{giulio.landolfi@unisalento.it; giulio.landolfi@le.infn.it}
\affiliation{$^a$ Institute of Industrial Science, 
University of Tokyo, Komaba, Meguro, Tokyo 153-8505, Japan \\
$^b$Department of Mathematics and Physics 'Ennio De Giorgi',  Universit\`a del Salento, 
via Arnesano, I-73100 Lecce, Italy \\
$^c$I.N.F.N. Sezione di Lecce, via Arnesano I-73100 Lecce, Italy}

\date{\today}

\begin{abstract}

We address the problem of integrating operator equations  concomitant 
 with the dynamics of non autonomous quantum systems by taking advantage of the use of time-dependent canonical transformations. In particular, we proceed to a discussion in regard to basic examples of one-dimensional non-autonomous dynamical systems enjoying the property that their Hamiltonian  
can be mapped through a time-dependent linear canonical transformation into an autonomous form, up to a time-dependent multiplicative
factor. The operator equations we process essentially  reproduce at the quantum level the classical integrability condition for these systems. 
Operator series form solutions in the Bender-Dunne basis of pseudo-differential operators for one dimensional quantum system are  sought for 
such equations. The derivation of generating functions for the coefficients involved in the \emph{minimal} representation of the series solutions 
to the operator equations under consideration is particularized. We also provide explicit form of operators that implement arbitrary linear 
transformations on the Bender-Dunne basis by  expressing  them in terms of  
the initial Weyl ordered basis elements. We then remark 
that the matching of the minimal solutions  obtained independently in the two basis, 
i.e. the basis prior and subsequent the action of  canonical
linear transformation,  is perfectly achieved  by retaining only the lowest order contribution in the expression of the transformed Bender-Dunne 
basis elements.

\end{abstract}

\pacs{  03.65 -w }

\maketitle

\section{Introduction}
\label{s0}
In classical mechanics integrability  is a fundamental concept  that has witnessed a broad range of developments in different directions (see, e.g., \cite{zakharov}-\cite{schwarzbach} and Refs. therein). Owing to the different nature of dynamical objects to be considered, and of 
diverse characteristics of the spaces thence implied, the attempt to transfer the same fabrics at the quantum level clearly suffers of some
inconsistencies.  Nonetheless, efforts to overcome difficulties are expected,
to improve  the understanding of  semiclassical regimes, and of classical to quantum transitions  (e.g., in respect to chaotic dynamics) or vice versa. To this, it makes particularly sense to consider the peculiar issues that may plague a satisfactory correspondence 
within formalism that better allow the realization of common views shared by classical and quantum mechanics. For instance,  approaches  based on phase-space distributions  appear well-suited to some purposes (among the others, see \cite{carmichael}-\cite{royer}).  
For a deeper insight it looks very reasonable to pay attention on independent conserved quantities and  algebras underlying an assigned problem, for instance  as  discussed  within the framework developed  after  the seminal papers \cite{barut,dothan} or as  currently pursued 
in consideration of the notion of \emph{superintegrability} (see e.g. (\cite{winternitz1,winternitz2}). 
Needless to say, to exploiting the results obtained in symmetry-based frameworks also supposes a profound comprehension of properties of the operators representing pairs of canonically conjugated quantum observables and of quantum implications of classical canonical transformations. 
 
In discussing integrable systems in classical mechanics,  the key role of canonical transformations cannot be ignored. Having proved to enable the gaining of general conceptual insights as well as the advantageous solving of a number of significant dynamical problems, their relevance is undoubted. 
Quantum canonical transformations, say the change of operator variables preserving the canonical commutation relations, may prove, in principle, similarly useful in connecting 
the wave equation solution for a system of interest to the solution of a simpler equation. 
Nevertheless, the topic enters in the discussion of quantum problems only limited, or with certain 
determined intents. In general, there are some reasons why this happens. 
In some cases, one may just prefer to derive approximate solutions because they are sufficient for arguing on a definite problem. 
In other cases, one may feel uncomfortable in dealing with transformations that only preserve the dynamical equations and not necessarily other ingredients one may like to retain {\em tout court}, such as the norm of states and the Hilbert space structure of the model.
But there are other primary obstacles. For instance, some caution is required in that the canonical transformation from position-momentum to action-angle coordinates 
are generally nonbijective (see \textit{e.g.} discussions in \cite{moshinsky, hakioglu}).
At the quantum level this unpleasant event is perceived as aggravated by the question raised about the uniqueness of the realization of operators obeying given commutation relations and  operator equations.  Moreover, the applicability of the standard correspondence principle, and of the outcoming uncertainty principles to arbitrary conjugate pairs generates controversial interpretations (see e.g. \cite{quantum phase in,quantum phase fin}). So, before trying to learn from the introduction of new observables, one may find desirable being in the position of adhering to a more satisfactory mathematical formulation of Quantum Mechanics that  excludes the appearance  of non-unitary features and ambiguities. 

Some of the above mentioned aspects are effectively grasped at once 
through a meaningful paradigmatic example. 
In a perspective slightly different from the standard Hamiltonian formulation, the problem of the integrability of a Hamiltonian system can be debated by putting some different emphasis on the time variable, in fact \footnote{We are aware that in the theory of dynamical systems and in mathematical physics integrability has acquired a meaning other than merely the success in the integration of equations \cite{zakharov}-\cite{schwarzbach}. 
In quantum mechanics, instead, the characterization of  exact or approximate  solutions to Schr\"{o}dinger equation has been the main issue so far, and only a minor focus has been put on 
other aspects, such to produce a more natural development from classical to quantum on 
geometric grounds \cite{wood}. 
On the other hand, problems we are going to deal with in this communication have a meaning close to that of Liouville integrability, but they will be addressed  having in mind
 the explicit construction of solutions as the definite object 
(while aspects concerning orbits/orbitals  will not be not mentioned).
For this, we relax the distinction between actual integration and integrable dynamics, and
we  use the term integrability somewhat roughly  to encompass both meanings.}.
For instance, for a one dimensional autonomous system whose Hamiltonian is 
\beq
H=\frac{p^2}{2m}+V(q) \,\, , 
\eeq
the problem of the integration of Hamilton equation of motion can be implemented by making use of the fact that the evolution of canonical coordinate $q$ can be given  implicitly via 
\beq
t=t_0+\int_{q_0}^q\frac{d\tilde{q}}{\sqrt{E-V(\tilde{q})}} \,\, ,
\label{eq int t}
\eeq
being $E$ the constant value taken by the Hamiltonian, i.e. the system's energy. 
Equation (\ref{eq int t}) clearly effects the possibility to describe the system dynamics, 
since a parameter (time or angle-like variable) that parametrizes continuous 
curves (or pieces of) in phase space can identify representative trajectories implicitly
preserving the existence of a dynamical group for time \footnote{The reader should be aware that the locution "dynamical group" here has a distinct meaning from  its use in the context inspired by papers \cite{barut,dothan}}.
In a sense, integration of the energy conservation law replaces the integration of one of the Hamilton equations.  One has, at variance, a focus on the relationship between the variable 
$F(q(t))=t +{\rm const} $, that is to say the variable whose Poisson brackets with the Hamiltonian $H$ equals to unity (i.e., the variable whose translations are the result of canonical transformations generated by the Hamiltonian). Armed with the map  between position and $F$, trajectories in phase space are likewise desumed from the Hamilton equations. 
In passing at the quantum level, one may perform a transposition of the argument for 
 Heisenberg equations of motion for position and momentum operators 
and  Heisenberg equation $[F,H]=i\hbar$
\footnote{For systems with more degrees of freedom, the argument logically requires the solving of equation of motion for  all but one the canonical variables. A simple example can be stated,  for instance, by means of  a central potential problem, in which case the coordinate $q$ in terms of which time is stated is nothing but the radial coordinate; see e.g. \cite{razavy radial}}. 
While it seems kind of an innocuous change of variables, dealing with the aspects connected with the new problem is actually an outstanding issue in standard formulation of Quantum Mechanics. It poses unavoidably  the question of the meaning of the classical-quantum correspondence and of 
the preservation of hermiticity and unitarity for operators entering in the framework
\cite{quantum phase in, quantum phase fin}, the doubts being supported by the famous Pauli's enbargo that there is no self-adjoint operator canonically conjugate to a Hamiltonian if the Hamiltonian spectrum is bounded from below. 
Remedies to ambiguities and inconsistencies have been valued, most significantly  by means of positive operator measures (see e.g. \cite{royer}) or confinement \cite{confined1,confined2},  
\footnote{It is worth to recall that 
for a free particle the Kijowski's  arrival distribution  and the Aharonov-Bohm time operator  have already been formally linked each to the other both by
means of the concept of positive operator valued measure in the
unconfined case \cite{giannitrapani}, and as limit of the distribution
one would obtain by restricting the position domain to a finite real
interval \cite{delgado}. }.  
Besides, within the phase-space distributions approach, a commonsensical  way to act
  can  be approved upon the suggestion that    
operators obtained by direct quantization (and introduction of an ordering rule) of classical phases
attribute a correct sharp phase to large amplitude localized states (see \cite{royer} and Refs. therein).
Hence, the formal solving of operator equations like the Heisenberg one above, $[F,H]=i\hbar$, 
always represents  an obliged road 
for comprehending to what extend corrective terms for observables may be extracted and 
how a picture should be modified in order to probe more general dynamical regimes and states
when deviation from classicality is considered. 

For non-autonomous systems the integrability issue  may be argued similarly, 
but some adaptations of the framework are needed.  
The question can be better elucidated from the point of view of the understanding of the time transformations required to fully preserve the structure of the equations (classical and quantum). Indeed, for general non-autonomous classical hamiltonian systems, symmetries are
associated with vector fields whose components along the time-derivative directions are not constant and depend on $(q,p,t)$. Mapping a non-
autonomous system into an autonomous one implies the accomplishment of a canonical transformation $(q,p,t) \rightarrow \left(Q(q,p,t),P(q,p,t),\tau(q,p,t) \right)$  into the time-extended phase-space $(q,p,t)$, in such a way that the symmetry vector field introduces a derivative with respect to the new time-like variable $\tau$. In practice, if this operation is acted out, a map is realized implying  the time-dependent Hamiltonian going into a dynamical invariant. At the classical level, one may expound this by saying that, being not fixed,  energy is 
not anymore the preferred observable for measuring features of orbits in phase-space, and transport of
dynamics of system states over them. (This is due to the correspondence Hamiltonian-energy for autonomous systems.) At the quantum level, 
this means that a basic stationary spectral quantum problem may be considered as opposite to the 
time-dependent Schr\"{o}dinger equation involving the  original Hamiltonian.  
Evolution equation through Poisson brackets or commutators are subject to adjustments accordingly. After doing so, the topic of their integrability may be assessed similarly  to the autonomous case.

Whichever the approach to the topic of quantum integrability and the peculiarities of Hamiltonians, 
operator equations entailing commutators needs definitely to be analyzed.
The task can be carried out by implementation of techniques accomplishing ordering of non-commuting operators (see e.g. \cite{B1,B2,Fan1,Fan2} and Refs therein).
Solving operator equation then essentially becomes into a matter of combinatorics, and interesting findings in this respect may be revealed in fact, see e.g. \cite{BDpoly, BGtime}.
A way to proceed is to resort to the technique proposed and discussed by Bender and Dunne in their enlightening paper \cite{B1},  based on the introduction of Weyl ordering (an ordering which holds interesting properties). It basically consists in looking for solution to an equation for an operator $\hat{F}(\hat{q},\hat{p})$
by finding its components in  a basis $\{ \hat{T}_{m,n}\} _{m,n\in \mathbb{Z} }$ 
whose elements $\hat{T}_{m,n}$ are constructed by superimposing Weyl ordering 
to products $\hat{p}^{m} \hat{q}^{n}$ of integer powers of position and momentum canonical operators. 
Hence the operator is sought  in the series form $\hat{F}=\sum_{m,n}\, f_{m,n} \hat{T}_{m,n}$,  the elements $\{T_{m,n}\}$'s being defined by applying the operator ordering prescription scheme
\beq
\hat{T}_{m,n}= \,\, :\hat{p}^m \hat{q}^n:_{Weyl}\,\,=
\frac{1}{2^{n}}\sum_{k=0}^{n}\binom{n}{k}\ \hat{q}^{k} \, \hat{p}^{m}\hat{q}^{n-k} =
\frac{1}{2^{m}}\sum_{k=0}^{m}\binom{m}{k}\ \hat{p}^{j} \, \hat{q}^{n}\hat{p}^{n-j}
\,\, 
\label{E12}
\eeq
where the first (second) ordered form has to be used for negative  momenta $p$ (position $q$) powers
\cite{B1}. Note that the binomials have to be expressed in terms of Gamma functions  whenever the power indexes $m$ or $n$ are negative. 
(For an analysis of properties of the Bender-Dunne basis $\{ \hat{T}_{m,n}\}$, see \cite{bunao}.)  
In order to lightening the  notation, and since no ambiguities can occur, the hat operator symbol $\,\hat{}\,$ shall be omitted hereinafter.

The operator equation considered in  \cite{B1,B2} was just the one finalizing an operator $F$ formally conjugated to the Hamiltonian, and solutions were given there for basic autonomous one-dimensional hamiltonian systems such as the particle undergoing the quadratic or the quartic potential. 
Evidently, and regardless the particular strategy possibly employed, whenever one attempts to solve the above equation in practice, one is generally faced with the question of the non-uniqueness of operators that can be formally identified as solutions of an operator equation (for  recent examples in this respect, see \cite{BGtime, BGC}). For this reason, 
the concept of \emph{minimal} solution has been introduced in  \cite{B1} as a viable criterion: it is 
the simplest among  the solutions that are compatible with the recursion relations implied  for  the expansion coefficients, {\em i.e.}  the solution  that minimize the number of nonzero coefficients. 
Remarkably, in \cite{B1} Bender and Dunne found that the minimal solution associated with the operator conjugated to the Hamiltonian in the harmonic oscillator case coincides with just the Weyl quantized form of the classical angle variable on the trajectory in the phase space.  
However, such a connection with the Weyl quantized form of the classical counterpart sounds incidental as the minimal solution strategy does not appear to carry this meaning in general. 
Indications to reconciling the minimal solution strategy introduced on pragmatic grounds with more authentic physical patterns have been elaborated in by resorting to the 
so-called inverse Liouville method.
 
When canonical transformations $\mathcal{C}$ enter in the treatment of a dynamical systems, 
other Bender-Dunne operator bases $\mathcal{T}_{m,n}$ are naturally defined in terms of the 
new pair of canonically conjugate operators.  Of course, the expansion of an observable is affected by a canonical transformation accordingly and expansion'  coefficients have to be properly modified in moving from one basis to the  other. 
Furthermore, if $\mathcal{C}$ explicitly depends on time, the property is inherited by components of basis elements  $\mathcal{T}_{m,n}$ in basis $\{T_{m,n}\}$.
One is faced with this situation whereby observables (e.g. solutions to some dynamical equations)
may be possibly sought in a new basis $\{\mathcal{T}_{m,n}\}$,  rather than  in the initial basis $\{T_{m,n}\}$, when there are hints that a canonical transformation is advantageous  in some respects to the system's investigation. One of such cases is the study of non-autonomous hamiltonian systems where extended generating functions can be identified that define canonical mappings $\mathcal{C}$ into time-independent hamiltonian problems. 
Remark that, working out the transformation, a fictitious time $\tau$ can be put forward 
that turns out to be correlated with the original  time parameter $t$ 
(sometimes referred as Poincaré -time transformation or as symplectic time rescaling method). In a sense, it acts as the proper time parameter whose invariance under translations is implied for the system described by the invariant  emerging from the original Hamiltonian thanks to the transformation, in the same manner as it happens for the initial 'physical' time $t$ and an autonomous Hamiltonian in its initial phase-space coordinates. Some attention may be so needed in moving consistently between the two hamiltonian formulations (prior ans subsequent the action of $\mathcal{C}$) while handling this time-diffeomorphism.

In the light of all the above,  in this communication we are aimed at exploring 
 issues raised and benefits possibly implied when time-dependent canonical transformations are 
resorted in the investigation of time-dependent operator equations for non-autonomous quantum  hamiltonian systems, such as the equation for their angle-action type operators. 
We shall pay attention to linear time-dependent canonical transformations for position and momentum variables, say $(q,p)\rightarrow (
Q,P)\equiv (A(t) \,q\,+D(t) \, p ,    B(t)\,p+C(t)\, q) $
 where $A,B,C$ and $D$ are real-valued functions of time $t$ obeying 
$A B-C D=1$ (only one-dimensional systems shall be considered), performed simultaneously to a time diffeomorphism. In the applications, we shall be particularly interested in the point-type limit $D=0$. 
This is the structure that emerges for the canonical transformation that maps time-dependent quadratic systems into time-independent ones.   
However, linear canonical transformations are of more general interest (see e.g. \cite{CLin}-\cite{CLfin} and \cite{hakioglu}) and resorting to them can be helpful  in other cases as well. 

The outline of the paper is as follows. In Section \ref{s1}, the time-dependent linear phase-space transformations in which we are interested  are introduced  
along with the accompanying unitary operator 
in order to point out  that, if a proper ansatz does hold, they enable to transform a non-autonomous system into an equivalent one which is autonomous. Once the main reason why we are interested  into  time-dependent  linear transformations is clarified, in Section \ref{s2} we consider their action on the Bender-Dunne basis of operators defined by equation (\ref{E12}). 
We widen partial results already found in this regard \cite{lohe} to the case of negative power of position (respectively, momentum) operators, \emph{i.e.} negative index $m$ (respectively, $n$) of the basis elements $T_{m,n}$ in  eq. (\ref{E12}), by explicitly expressing the result in manifestly ordered normal and Weyl forms. 
Section \ref{classical vs quantum} is aimed at providing an 
immediate intuition on the structural difference between the formal position-momentum expansions of Weyl-quantized classical solutions to the integrability condition for one-dimensional hamiltonian systems and those of quantum solutions (including minimal ones). 
The succeeding sections are devoted  to the processing of the integrability condition in some concrete examples, such as non-autonomous one-dimensional quantum systems undergoing the linear, the quadratic and the quartic potentials.
All the  cases are treated dwelling on details for enabling the reader to get more acquainted with the  framework. This should serve  particularly while applying it  further for applications and developments in respect to to equations and operators that appear to be defined, or of interest, only at the quantum level.  Finally, Appendix A summarizes the inverse Liouville method.

\section{Linear phase-space transformations }
\label{s1}

Although not very customary (even in the classical setting), the treatment of a non-autonomous system with Hamiltonian $H_{na}(p,q,t)$  may be  concerned ultimately with contact transformation in the extended phase-space in such a way that the new Hamiltonian is actually autonomous. In practice, this means to deal with extended canonical transformations $(q,p,t)\rightarrow (Q,P,\tau)$
that mix position and momentum (coordinates or operators, reliant on the dynamical regime) 
through a time dependent  phase-space map implemented by a redefinition of the time variable 
see e.g. \cite{struckmeier, johns} (for a discussion treating with  time-dependent quadratic systems, see also \cite{B4}).
As we already said,  the basic $Q,P$ structure that enters in the analysis of the non-autonomous Hamiltonians in which we are interested is linear in the original canonical pair $(q,p)$. Below, we shall 
clarify  the purposes of use. 

\noindent Let us consider the one degree of freedom quantum time-dependent Hamiltonian
\begin{equation}
H=\frac{p^2}{2m(t)}+V(q,t) \,\, .
\label{a2}
\end{equation}
In general, a  unitary transformation $U(q,p,t)$ maps the original Hamiltonian (\ref{a2}) into a  new 
 operator $H^\prime$, according to 
\beq
H'=U^{-1} H U-i U^{-1} \partial_t U \,\, . 
\eeq
Attempts can be made  to map the original Hamiltonian (\ref{a2}) into a new Hamiltonian taking a convenient form. 
Let us focus, in particular, on extended canonical transformations 
that are linear in  position-momentum variables, and are defined according to
\beq
Q=A(t) \,q\,\,,\qquad P=B(t) \, p+C(t)\, q \,\,  , \qquad {\rm with} \quad A(t)\,B(t)=1\,\, .
\label{a1}
\eeq
It is worth to remark that this transformation is unitary and consists of a two-step action in phase-space: a time-dependent dilatation of the configuration space, 
 and a momentum translation that depends on the canonical position linearly. 
The composition of such kind of two actions can be identified through 
unitary operators of the form $U_2= {\rm exp} [ i \Delta_2(t) \,\, (pq+qp)]$ 
and $U_1={\rm exp} [ \Delta_1(t)\,\, q^2]$, 
respectively. Importantly the composition of  two time-dependent actions of this type 
can marginalize the role of time in non-autonomous hamiltonian quadratic systems, 
in the sense that solving the Schr\"odinger equation basically reconciles with time-independent spectral problems  in the new canonical coordinates (see, e.g., discussions in \cite{B6, B4}). 
To be precise, if the Hamiltonian is given in the form 
\beq
H=\frac{p^2}{2m(t)}+\frac{m(t) \,\,  \omega(t)^2}{2} q^2  \,\, ,
\label{H quadratic}
\eeq  
this happens provided that \cite{B6} 
\begin{eqnarray}
U_1&=& {\rm Exp} \left\{- i\frac{\dot{\rho}(t)\rho(t)m(t)}{2}q^2\right\}\,\, ,\nonumber \\
U_2&=& {\rm Exp}\left\{ \frac{i}{2}\log\left[\rho(t)\right](pq+qp)\right\}\,\, ,
\label{a3}
\end{eqnarray}
where $\rho(t)$ depends on the potential frequency-type function $\omega$ and mass-type function $m$ through  the differential equation 
\beq
m(t)\ddot{\rho}(t)+\dot{m}(t)\dot{\rho}(t) +m(t) \, \omega(t)^2 \, \rho(t)=\frac{1}{m(t)\rho(t)^3}\,\, ,
\label{a4 ermakov}
\eeq
(which is more often presented as an Ermakov equation \cite{ermakov, ibragimov} 
for the function $\sigma=\kappa \sqrt{m} \rho$ with $\kappa={\rm const}$, see e.g. \cite{glprd, cessa} and refs. therein).
So, the transformation (\ref{a1}) is applied with time-dependent coefficients  
\beq
A(t)=\rho(r)^{-1},\quad B(t)=\rho(r), \quad C(t)=-m(t)\dot{\rho}(t) \,\, ,
\label{a7}
\eeq
to give $H'=(2m \rho^2)^{-1} (P^2+Q^2)$.
The question arises to this point about to what extend the transformation can also be tried out on other systems   with the scope to endeavor to a new handy Hamiltonian.  The simplest case would be   
if it will take the form of a time independent Hamiltonian  multiplied by a time dependent factor. 
This convenient chance is not to be excluded {\em a priori}, in fact. 
Precisely, we remark here that whenever the condition 
\beq
m(t)\rho^2(t) V(\rho(t) q,t)=W(q)
\label{a5}
\eeq
holds true for a non-quadratic time-dependent potential $V$ entering an Hamiltonian operator $H$ defined as in (\ref{a2}), the net action on the Hamiltonian  of the unitary transformation 
$U=U_1U_2$, with $U_{1,2}$ given by (\ref{a3})-(\ref{a4}),  results into the new Hamiltonian operator
\beq
H'=\frac{1}{m(t)\rho^2(t)} \, H_0' \,\, , \qquad \qquad {\rm where} \qquad H_0'=\frac{P^2+Q^2}{2} +W(Q)\,\,
\label{a6}
\eeq
 provided that 
the scaling function $\rho$ and the mass-type function $m(t)$ satisfy the differential equation
\beq
m(t)\ddot{\rho}(t)+\dot{m}(t)\dot{\rho}(t) =\frac{1}{m(t)\rho(t)^3}\,\, .
\label{a4}
\eeq
In fact, for potentials other than the quadratic one as in Hamiltonian (\ref{H quadratic}), it turns out that, demanding 
the final Hamiltonian to be time-diffeomorphic to an autonomous operator, the single condition 
(\ref{a4 ermakov})  is actually replaced by two distinct conditions: the algebraic condition (\ref{a5}) 
selecting the class of potentials for which the request can be successfully fulfilled, 
and the differential equation (\ref{a4}) for the identification of the proper scaling function $\rho$ 
giving rise to the desired canonical transformation (as a matter of fact, the two conditions merge in the case of quadratic potentials (\ref{H quadratic}) because both appear in connection with coefficients of quadratic position terms $Q^2$ generated via the transformation).
In a slightly different perspective, it can be said that a dynamical invariant   $I =U H_{0}' U^{-1}$  can be obtained for the system described by the Hamiltonian (\ref{a2}).  In fact, suitable conditions are so identified for the coefficients in Eq. 
(\ref{a1}) such that the change of variables $ (q,p,t) \leftrightarrow\left(Q ,P ,\tau \right)$   enables to express  the dynamical variable $I $ as a 
stationary one with respect to $(Q,P)$ phase-space coordinate (i.e., $I(p,q,t)\rightarrow I(P,Q)$). In particular, this is achieved by setting 
coefficients as in (\ref{a7}) and by performing the change of  time variable via
\beq
\tau=\frac{1}{2}\int_{t_{0}}^{t}\frac{1}{m(t')\rho(t')^{2}}dt' \,\, .
\label{a8}
\eeq

\section{Weyl ordered series expansion of the operator $\frac{1}{p^{m}}q^{n}$
under linear phase-space transformations }
\label{s2}

A canonical transformation $\mathcal{C}$ on phase-space  naturally induces a transformation  on dynamical  variables on the same space. If the transformation is time-dependent, the immediate quantum analogous of a classical dynamical variable is realized through time-dependent maps acting on the individual operators entering its definition. In particular, if the canonical transformation $\mathcal{C}$ is applied to Bender-Dunne basis elements $T_{m,n}$, then Weyl ordered bases with elements   $\mathcal{T}_{m,n}=:
\mathcal{C} (T_{m,n}):_{Weyl}$  may be defined at different times.  In this section, we will determine the explicit representation of operators  $\mathcal{T}_{m,n}$ in the basis $\{T_{m,n}\}$ once a linear  transformation mapping is applied to position and momentum. 
We consider first the most general case where
\beq
Q=A\,q+D\,p,\qquad P=B\,p+C\,q\,\, ,
\label{g1}
\eeq
(canonical iff $A B-CD=1$) and we will then consider the canonical transformation stemming from (\ref{a1}) with time dependent coefficients given in (\ref{a7}), viz. 
\begin{equation}
Q= \frac{q}{\rho(t)}\,\, ,\qquad 
P=\rho(t) \, p -m(t)\, \dot{\rho}(t) \,q \,\, .
\label{E1}
\end{equation}
To achieve our goal, we adopt a strategy based on step-by-step calculations as follows. Introduce the Bender-Dunne basis associated with the new canonical operator pair $Q$ and $P$, that is the operator basis made of the Weyl ordered elements
\footnote{Formal questions  can be obviously raised in respect to the definition and the action 
of position and momentum operators and of their inverse, which in turn address to the properties of canonically conjugated operators. Implications can be discussed similarly  to the case of operators $T_{m,n}$; see, for instance,  investigations \cite{B3,B7} for possible consequences of confinement. }: 
\beq
\cT_{-m,n}=:P^{-m} Q^n:_{Weyl} \, =\frac{1}{2^{n}}\sum_{k=0}^{n}\binom{n}{k}\ Q^{k}P^{-m}Q^{n-k}
\label{g3}
\eeq
We are interested here in finding a closed form for the operators (\ref{g3}) in terms of initial variables 
$q,\ p$ and $t$. Once the  canonical transformation (\ref{g1}) is given, the $\cT_{-m,n}$'s can be seen as depending on the initial variables $(q,\ p,\ t)$, indeed. In particular, the $\cT_{-m,n}$'s can be given in terms of the Weyl ordered expansions of products between powers of powers of $q$ and $p$. This is straightforward if $C=0$, in that the canonical transformation merely implies a time-dependent scaling of position and momentum, and accordingly $\cT_{-m,n} \rightarrow b^{m+n} T_{-m,n}$. 
But if $C\neq 0$, the relationship between the two basis $\{\cT_{-m,n}\}$ and $\{ T_{-j,k}\}$ is obviously not that trivial.  Indeed, while
the positive powers of the operator $Q$  can be easily expressed in terms of $(q,p,t)$ as
\beq
Q^\alpha=(D\,p+A\,q)^\alpha=\sum_{s=0}^\alpha\binom{\alpha}{s}\frac{D^{\alpha-s}A^s}{2^{\alpha-s}}\sum_{u=0}^{\alpha-s}\binom{\alpha-s}{u}p^u\,q^s\,p^{\alpha-s-u} \,\, 
\label{g4}
\eeq
where in (\ref{g3}) $\alpha=k$ and $\alpha=n-k$,  
the situation is more delicate for the negative power of the operator $P$, that we can write as
\beq
P^{-m} =C^{-m}p^{-m}\,\, \frac{1}{1+\left(\frac{C}{B}\right)^m\sum_{j=0}^{m-1} D_{m,j}T_{j,m-j}p^{-m}}
\,\, .\label{g5}
 \eeq
We can then render  operators (\ref{g3}) in the form
\beqa
\cT_{-m,n}^{\cC}&=&\frac{D^n}{4^n\,C^m} \sum_{k=0}^n\binom{n}{k}\sum_{s=0}^k\binom{k}{s}\frac{2^s A^s}{D^s}\sum_{u=0}^{k-s}\binom{k-s}{u}\sum_{t=0}^{n-k}\frac{2^t A^t}{D^t}\sum_{v=0}^{n-k-t}\binom{n-k-t}{v} \,\,\, \times\nonumber \\
& & \qquad \qquad \qquad\times \,\,\, p^u\,q^s\,p^{k-s-u-m}\left(1+\frac{C^m}{B^m}\sum_{j=0}^{m-1} D_{m,j}T_{j,m-j}p^{-m}\right)^{-1}p^v\,q^t\,p^{n-k-t-v}.
\label{g6}
\eeqa
(We shall use the upperscript $^{\cC}$ to point out that the canonical relationship (\ref{g5}) has been applied to express operators in the $(q,p)$-representation). By resorting to the series expansion $(1+\Delta_m)^{-1}=\sum_{r=0}^\infty (-\Delta_m)^r$ for the operator 
\beq 
\Delta_m=\frac{C^m}{B^m}\sum_{j=0}^{m-1} D_{m,j}T_{j,m-j}p^{-m},
\label{g7}
\eeq 
we can then express (\ref{g6}) as
\beq
\cT_{-m,n}^{\cC}=\frac{D^n}{4^n\,C^m} \sum_{r=0}^\infty (-1)^r \sum_{k=0}^n\binom{n}{k}\sum_{s=0}^k\binom{k}{s}\frac{2^sA^s}{D^s}\sum_{u=0}^{k-s}\binom{k-s}{u}\sum_{t=0}^{n-k}\frac{2^t A^t}{D^t}\sum_{v=0}^{n-k-t}\binom{n-k-t}{v} \mathcal{O}^r_{k,m,n,s,t,u,v},
\label{g8}
\eeq
 where
 \beq
 \mathcal{O}^r_{k,m,n,s,t,u,v}=p^u\,q^s\,p^{k-s-u-m}\Delta_m^r\,p^v\,q^t\,p^{n-k-t-v}.
 \label{g9}
 \eeq
At this stage, the operators $\cT_{-m,n}^{\cC}$ are given by successive actions of different products 
of powers of the initial noncommuting canonical operators $q$ and $p$.  
 Formal simplification of the above operator series can be realized for making as clear as possible the action of the operators $q$ and $p$. The most effective way for proceeding is to introduce normal ordering in each the operators (\ref{g9}), by making use of the formula
 
\beq
q^kp^{\ell}=\sum_{\alpha=0}^k B_{\alpha,k,\ell}\,p^{\ell-\alpha}q^{k-\alpha} \,\, ,
\eeq
with 
\beq 
B_{\alpha,k,\ell}=\frac{(-i\hbar)^\alpha \,\,  \Gamma(k+1)\,\, \Gamma(\ell+1)}{\alpha!\,\, \Gamma(k-\alpha+1)\,\, \Gamma(\ell-\alpha+1)},\quad (k\geq0,\,\ell>0)
\label{g10}
\eeq
for $\ell>0$, otherwise
\beq
B_{\alpha,k,\ell}=\frac{(-i\hbar)^\alpha \,\, \Gamma(k+1)\,\, \Gamma(\alpha-\ell)}{\alpha!\,\, \Gamma(k-\alpha+1)\,\, \Gamma(-\ell)},\quad (k\geq0,\,\ell<0).
\label{g11}
\eeq
 By virtue of this, Equation (\ref{g9}) can be afterward expressed  as
\beq
 \mathcal{O}^r_{k,m,n,s,t,u,v}=\frac{C^{mr}}{B^{mr}}\underbrace{\sum_{j_i=0}^{m-1}D_{m,j_i}\,\sum_{\alpha_i=0}^{\delta_i} B_{\alpha_i,\delta_i,\gamma_i}}_{i=1,\,2,\,\dots,\,r+1}\,\,p^{n-(r+1)m+J_r-A_r-s-t}\,q^{rm-J_r-A_r+s+t}
 \eeq
where
\beq
J_r=\sum_{\ell=1}^r j_\ell,\quad A_r=\sum_{\ell=0}^{r+1}\alpha_\ell,
\eeq
and
\beq
\delta_i = \begin{cases} t &\mbox{if  } i = 0 \\ 
m-j_i & \mbox{if  }  1\leq i\leq r\\
s & \mbox{if  }  i=r+1 \end{cases},  \quad \quad
\gamma_i = \begin{cases} n-k-t-v &\mbox{if  } i = 0 \\ 
n-i\,m-k-t+J_i-A_{i-2}-\beta_i & \mbox{if  }  1\leq i\leq r\\
n-i\,m-s-t-u+J_{i-1}-A_{i-1} & \mbox{if  }  i=r+1 \end{cases} 
\eeq
To invoke back the $T_{m,n}$ basis defined through equation (\ref{E12}),
we will use the relation  
\beq
p^aq^b=T_{a,0}T_{0,b}=\sum_{c=0}^{\infty}\sum_{d=0}^{j}\,F_{a,b,c,d}\,\, T_{a-c, b-c}
\label{g12a}
\eeq
with
\beq
F_{a,b,c,d}=\left[ \frac{1}{c!} \left(\frac{i}{2}\right)^c\right]
\left[ \frac{c!}{  (c-d)! \, d!} \right]
\left[\frac{\Gamma(a+1) \,\, \Gamma(b+1)}{\Gamma(a+1-d) \,\Gamma(b+1-d) \, \Gamma^2(d-2c)}\right] \,\, , 
\label{g12b}
\eeq 
that can be simplified according to
\beq
p^{a}q^b=T_{a,0}T_{0,b}=\sum_{c=0}^{b}\left(\frac{-i \hbar}{2}\right)^c\frac{\Gamma(1-a) \,\, \Gamma(1+b)}{\Gamma(1-a-j) \,\, \Gamma(1+b-j)}T_{a-c,b-c} \,\, 
 \label{g13}
\eeq
whenever the powers of $p$ are always non-positive and the powers of $q$ are always non-negative.
We obtain therefore the final result for formula (\ref{g9}): 
 \beq
 \mathcal{O}^r_{k,m,n,s,t,u,v}=\sum_{c=0}^\infty\sum_{d=0}^j\underbrace{\sum_{\alpha_i=0}^{\delta_i} B_{\alpha_i,\delta_i,\gamma_i}}_{i=1,\,2,\,\dots,\,r+1}\,F_{a,b,c,d}\,\, T_{a-c, b-c} \,\, ,
 \eeq
 where $a=n-(r+1)m+J_r-A_r-s-t$ and $b=rm-J_r-A_r+s+t$. This completes the derivation of how Weyl-ordered Bender-Dunne basis elements transform under linear transformations $(\ref{g1})$.

\subsection{The subclass of canonical transformations (\ref{E1}).}

Let us now consider the canonical transformation (\ref{a1}) with time dependent coefficients given in (\ref{a7}), i.e. Eq. (\ref{E1}). 
In this case the general structure for the operators $\mathcal{T}_{m,n}^{\cC}$ simplifies to 
\beq
\cT_{-m,n}^{\cC}=\frac{1}{2^n\,\rho^{(m+n)}} \sum_{r=0}^\infty (-1)^r \sum_{k=0}^n\binom{n}{k} \mathcal{O}^r_{k,m,n} \,\, ,
\label{E2}
\eeq
 where
 \beq
 \mathcal{O}^r_{k,m,n}=\left(\frac{-m \dot{\rho}}{\rho}\right) ^{m r} \,q^k\,p^{-m}\,\left[\sum_{j=0}^{m-1}D_{m,j}\,T_{j,m-j}\,p^{-m}\right]^r\,q^{n-k} \,\, .
 \eeq
By virtue of this, after some manipulations, we can write a closed form expression for each operator  $\mathcal{O}^r_{k,m,n}$ :

\beq
\mathcal{O}^r_{k,m,n}=\frac{1}{2^n}\left(\frac{-m \dot{\rho}}{\rho}\right) ^{m r} \,\sum_{k=0}^n\binom{n}{k}\underbrace{\sum_{j_i=0}^{m-1}\frac{D_{m,j_i}}{2^{j_i}}\sum_{\beta_i=0}^{j_i}\binom{j_i}{\beta_i}\sum_{\alpha_i=0}^{\delta_i}B_{\alpha_i,\delta_i,\gamma_i}}_{i=1,\,2,\,\dots,\,r+1.}p^{-(r+1)m+d_r}q^{n+rm-s_r} \,\, ,
\label{E9}
\eeq
where
\beq
d_r=\sum_{\ell=1}^{r+1}(j_\ell-\alpha_\ell),\quad s_r=\sum_{\ell=1}^{r+1}(j_\ell+\alpha_\ell), \,\,  
\label{E10}
\eeq

\beq
\delta_i=m-j_i,\quad (i=1,\,2\,\dots\,r),\quad \delta_{r+1}=k \,\, 
\label{E11}
\eeq
and 
\beq
\gamma_i=\sum_{\ell=1}^i j_\ell-\sum_{\ell=1}^{i-1}\alpha_\ell-\beta_i-i\,m
=d_{i-2}+j_i-\beta_i-i\,m 
\label{E12a}
\eeq
(we have   assumed that $\sum_{\alpha=n_0}^n=0$ if $n<n_0$ and $\beta_{r+1}=j_{r+1}=0$). 
Hence, taking into account (\ref{g13}), from equation (\ref{E9}) we obtain

\beq
\mathcal{O}^r_{k,m,n}=\frac{1}{2^n}
\left(\frac{-m \dot{\rho}}{\rho}\right) ^{m r} \,\sum_{k=0}^n\binom{n}{k}\,\underbrace{\sum_{j_i=0}^{m-1}\frac{D_{m,j_i}}{2^{j_i}}\sum_{\beta_i=0}^{j_i}\binom{j_i}{\beta_i}\sum_{\alpha_i=0}^{\delta_i}B_{\alpha_i,\delta_i,\gamma_i}}_{i=1,\,2,\,\dots,\,r+1.}\sum_{c=0}^b\left(\frac{-i \hbar}{2}\right)^c\frac{\Gamma(1-a)\,\,\Gamma(1+b)}{\Gamma(1-a-j)\,\,\Gamma(1+b-j)}T_{a-c,b-c}.
\label{E15}
\eeq
with  $a=-(r+1)m+d_r,b=n+rm-s_r$. 
Since $a$ is always negative, we can use the identity
$$\frac{\Gamma(1-a)}{\Gamma(1-a-j)}=(-1)^j\frac{(-a-1+j)!}{(-a-1)!}, \quad a<0$$
and perform the change of variable $a \rightarrow-a$ in (\ref{E15}), that becomes
\beq
\mathcal{O}^r_{k,m,n}=\frac{1}{2^n}
\frac{\left(-m \dot{\rho}\right)^{mr}}{\rho ^{m(r+1)}}
\,\sum_{k=0}^n\binom{n}{k}\,\underbrace{\sum_{j_i=0}^{m-1}\frac{D_{m,j_i}}{2^{j_i}}\sum_{\beta_i=0}^{j_i}\binom{j_i}{\beta_i}\sum_{\alpha_i=0}^{\delta_i}B_{\alpha_i,\delta_i,\gamma_i}}_{i=1,\,2,\,\dots,\,r+1.}\sum_{c=0}^b\left(\frac{i \hbar}{2}\right)^c
\frac{(a+j-1)!b!}{(b-j)!(a-1)!}
T_{-a-c,b-c} \,\, , 
\label{E16}
\eeq
where now $a=(r+1)\,m-d_r$, $b = -\,n - r\,m + s_r $. 
In turn, it follows that the operators  $\cT_{-m,n}^{\cC}$ in (\ref{E2}) 
are expressed in terms of the Bender-Dunne basis operators (\ref{E12}) by means of the formula 
\beq
\cT_{m,n}^{\cC}=\frac{1}{2^n}\frac{\left(-m \dot{\rho}\right)^{mr}}{\rho(t)^{n+m(r+1)}}\sum_{r=0}^\infty(-1)^r\sum_{k=0}^n\binom{n}{k}\,\underbrace{\sum_{j_i=0}^{m-1}\frac{D_{m,j_i}}{2^{j_i}}\sum_{\beta_i=0}^{j_i}\binom{j_i}{\beta_i}\sum_{\alpha_i=0}^{\delta_i}B_{\alpha_i,\delta_i,\gamma_i}}_{i=1,\,2,\,\dots,\,r+1.}\sum_{c=0}^b\left(\frac{i \hbar}{2}\right)^c
\frac{(a+j-1)!b!}{(b-j)!(a-1)!}
T_{-a-c,b-c} \,\, , 
\label{E17}
\eeq
 Formula (\ref{E17}) has a rather different structure as compared to the one attributed to the polynomials 
 associated with the product of positive powers of position and momentum operators 
subjected to  linear  transformations \cite{lohe}. 
In particular,  it is represented by an infinite sum of operator terms.

 \subsection{The action of canonical transformation (\ref{E1}) on operators $\cT_{-n,n}$}
Having in mind the first results obtained making use of formal expansions in the Weyl-ordered basis (\ref{E12}), i.e. the discussion by Bender and Dunne in \cite{B1} dealing with the very fundamental case of the  harmonic oscillator, and remembering  how a link with its natural time-dependent generalization is established by means of linear canonical transformations \cite{B4}, in this Subsection we shall confine the attention on the particular case $m=n$. Furthermore, since the \emph{minimal} solution for the observable conjugated to the quadratic Hamiltonian coincide with the Weyl-quantized form of the solution to the classical problem, for our scope suffices the selection of the classical limit of equation (\ref{E17}). 
For this particular case $m=n$, the zero-th term order in $\hbar$ corresponds to $\alpha_j=0$ ($j=1,2,\dots r$) and $c=0$ in Eq.(\ref{E16}). 
The criterion significantly simplifies the formula for the operator $\cT_{-n,n}^{\cC}$ 
by providing
\beq
\cT_{-n,n}^{\cC}=\rho ^{-2n}T_{-n,n}+\rho ^{-2n}\,\,\sum_{r=1}^\infty 
(-1)^r \left( \frac{-m \dot{\rho}}{\rho}\right)^{nr}
\,\,\, \underbrace{\sum_{j_i=0}^{n-1}\binom{n}{j_i}\left(\frac{-\rho}{m \dot{\rho}}\right)^{j_i}}_{r-times,\,\, 1\leq i\leq r}\,T_{-(r+1)n+\sum_{i=1}^r j_i,\,(r+1)n-\sum_{i=1}^r j_i} \,\, ,
\label{E19}
\eeq
a result that  can be effectively made plainer in the form  
\beq
\cT_{-n,n}^{\cC}=\rho ^{-2n}T_{-n,n}+
\frac{1}{\left(m\rho  \dot{\rho}\right)^n}\sum_{r=n+1}^\infty (-1)^r\binom{r-1}{n-1} 
\left(\frac{-m \dot{\rho}}{\rho }\right)^rT_{-r,r} \,\, , 
\label{E20}
\eeq
where the $r=0$ term has been isolated for the sake of a more direct comparison with the case prior the transformation.  Actually, it is also instructive to point out that
 the series (\ref{E19})-(\ref{E20}) can be also  manipulated to finally obtain
\begin{eqnarray}
\cT_{-n,n}^{\cC}=\frac{ X^n}{2\rho^{2n}}\left[ 1-\frac{m \dot{\rho}}{\rho}X\right]^{-n} +\,\,\frac{ Y^n}{2\rho^{2n}}\left[1-\frac{m \dot{\rho}}{\rho}Y \right]^{-n} \,\, , 
\label{E22}
\end{eqnarray}
being  $X$ and $Y$  the two operators
\beq
X=\frac{1}{p}\, q\,\, ,\qquad Y=q\, \frac{1}{p} \,\, .
\eeq
Precisely, the latter form can be obtained by  making essential use of the worthwhile identities
\beq
T_{-r,r}=\frac{1}{2}\left(\frac{1}{p}q \right)^r +\frac{1}{2}\left(q \frac{1}{p} \right) ^r=\frac{1}{2}X^r+\frac{1}{2}Y^r \,\, ,
\eeq
and 
\beq
\sum_{r=n+1}^\infty  (-1)^r\binom{r-1}{n-1}z^r=z^n\left(-1+(1-z)^{-n}\right) \,\, .  
\eeq
Equation (\ref{E22}) may in fact effectively work as  an operating expedient in some investigations. 

\section{Classical vs. quantum position and momentum expansions of observables
\label{classical vs quantum}}

In next Sections,  some examples will be given as of the application of Weyl-ordered  basis {\em a l\`a} Bender-Dunne  to basic problems pertinent the integrability of quantum non-autonomous systems.
Before doing so, however, we would like to make some remarks on the reliability of quantization and dequantization procedures in connection to a given hamiltonian problem.  
In fact, direct quantization of classical solutions in phase space generally fails (with exceptions whose existence will be intelligible later) in representing the solution to the quantum  version of the same equations obtained 
on algebraic grounds via the substitution of Poisson brackets with commutators. A quantum operator equations has to be investigated {\em per se}, even if we already have a solution for its classical relative. 
In translating a classical dynamical equation into a quantum one, there is an issue in respect to the proper transfer of algebraic features into 
the solutions. The root cause is that at the classical level there are no ordering rules, as opposite to the quantum level: the difference equation
one would obtain for expansion coefficients in a classical  basis of a classical dynamical object obeying an algebraic condition (in the Poisson brackets algebra sense) commonly differs from the difference equation one would obtain for expansion coefficients in a quantum analogous of the same basis for the observable required to obey the  condition after replacing  Poisson brackets through commutators.
If we have a classical observable represented in one basis (like 
the set of products of position and momentum powers $t_{m,n}=p^m q^n$, associated with Laurent-type expansions), relations between
coefficients are structurally simpler because the  Poisson-brackets formalism introduces algebraic aspects that only concern scalar functions
and their derivatives, i.e. commutative objects.
In different words, while the direct dequantization of operator equations \cite{maslov} and theirs solutions ends up into the forms arising for the
solution to the classical problem, the converse is not true: direct quantization of a solution to a classical observable equation  does not capture all the features implied by the solution of the  quantum version of the dynamical equation. For arbitrary dynamical variable the aspect is an acute, in the meaning which can be desumed for the simple discussion below. For illustrating effectively the point, it is enough
   to focus on the observable canonically conjugated to the Hamiltonian in the simple case of an autonomous system whose quantum (classical) dynamics is conducted by a power-law potential, namely by the quantum (classical) Hamiltonian 
\beq
H_L=\frac{p^2}{2}+\frac{q^L}{L}
\label{H L}
\eeq
where $L$ is a positive integer. The discussion can be immediately adapted to other observables and to the class of non-autonomous analytical potentials, though. Let us consider therefore
the formal expansion in the Bender-Dunne basis 
\beq
\Theta_L=\sum_{m,n}\alpha_{m,n}T_{m,n} \,\,
\eeq
for the operator $\Theta_L$ canonically conjugate to $H_L$ through the operator equation 
$[ \Theta_L,\,H_L]=i\,\hbar$. From the algebra of the $T_{m,n}$ operators
\begin{align}
\left[T_{m,n},T_{r,s}\right]&=2\sum_{j=0}^\infty \frac{1}{(2j+1)!} \left(\frac{i\hbar}{2}\right)^{2j+1} 
\,\,\sum_{l=0}^{2j+1} (-1)^l  \,\binom{2j+1}{l} 
 \quad \times \nonumber \\ 
&
\qquad \qquad\times \qquad \frac{\Gamma(m+1) \Gamma(n+1)\Gamma(r+1)
\Gamma(s+1)}{\Gamma(m-l+1)\Gamma(n+l-2j)\Gamma(r+l-2j)\Gamma(s-l+1)}
\,\, T_{m+r-2j-1,n+s-2j-1}
 \nonumber\\ 
&= i\hbar (n\, r -m\, s) \,\, T_{m+r-1, n+s-1} 
+2\sum_{j=1}^\infty \left(
\frac{i\hbar }{2}\right)^{2j+1}\frac{  \Gamma(n+1)\Gamma(r+1)}{(2j+1)!\, 
\Gamma(n-2j)\Gamma(r-2j) } \quad \times
 \nonumber\\ 
& \qquad \qquad \qquad \qquad \times \quad  
\,\, _3F_2[-1-2 j,-m,-s;-2 j+n,-2 j+r;1]
\,\, T_{m+r-2j-1,n+s-2j-1}
\label{algebra T}
\end{align} 
we know that
\beqa
[T_{m,n},\,p^2] &=& 2\,i\,\hbar\,n\,T_{m+1,n-1}\,\, , \\ \nonumber 
[T_{m,n}, q^L] &=& -i\,\hbar\,L\,m\,T_{m-1,n+L-1} -i\,\hbar\sum_{j=1}^{\lfloor(L-1)/2 \rfloor}\hbar^{2j}A_{m,j,L}\,T_{m-2j-1,\,n+L-2j-1}\,\, ,
\eeqa
where the nonvanishing coefficients $A_{m,j,L}$ can be inferred from Eq. (\ref{algebra T}) with $r=0$ and $s=L$, 
\beq
A_{m,j,L}=\frac{(-1)^j\,\Gamma(m+1)\,\Gamma(L+1)}{4^j\,(2j+1)!\,\Gamma(m-2j)\,\Gamma(L-2j)}
\,\, ,
\eeq
So the quantum canonical conjugation relation $[\Theta_L,\,H_L]=i\,\hbar$  can be complied with if 
coefficients characterizing the realization of the observable $\Theta_L$ fulfill the recursion equation
\beq
(n+1)\alpha_{m-1,n+1}-(m+1)\alpha_{m+1,n-L+1}-\frac{1}{L}\sum_{j=1}^{\lfloor(L-1)/2 \rfloor}\hbar^{2j}\,\alpha_{m+2j+1,n-L+2j+1}\,A_{m+2j+1,n-L+2j+1,L,j}=\delta_{m,n}\delta_{n,0}
\label{recursion equation quantum}
\eeq
(we have used that $1=\sum_{m,n}\delta_{m,0}\delta_{n,0}T_{m,n}$).
If  we  consider instead the problem classically, and resort to the use of the classical counterpart  
of the Bender-Dunne basis in phase space, the expansion in the $(p,q)$ coordinates
pair 
\beq
\Theta_L^c=\sum_{m,n}\alpha^c_{m,n} \, \, t_{m,n}=\sum_{m,n}\alpha^c_{m,n}\,\, p^m\,q^n,
\eeq
would be introduced for the variable $\Theta_L^c$ conjugated to the classical form of Hamiltonian
$H_L$. This successively means that the classical condition of canonic pairing 
involving the Poisson brackets $\lbrace \Theta_L^c\, H^c_L  \rbrace= 
  p\,\partial_q\Theta_L^c-q^{L-1}\,\partial_p\Theta_L^c=1 $ generates the difference equation 
\beq
(n+1)\, \alpha^c_{m-1,n+1}-(m+1)\, \alpha^c_{m+1,n-s+1}=\delta_{m,0}\delta_{n,0} 
\label{recursion equation classical}
\eeq
for the coefficients $\alpha^c_{m,n}$.
Hence, direct quantization of the dynamical variable $\Theta_L^c$ through the mere substitution of the basis elements $t_{m,n}$ with their quantum analogous $T_{m,n}$ is not enough to comprise additional  features of quantum analogue of the same dynamical problem due to the  quantum ordering. In fact, while dequantization of an observable solving an assigned quantum equation naturally ends up in its classical counterpart obtained by solving the same formal equation after the replacement of commutators through Poisson brackets, something may be lost  when  the converse procedure of  straight quantization of a classical observable is performed.
 
\section{Examples. I. Time-dependent quadratic hamiltonian systems
 \label{s3}}

As we anticipated, in the present and the next Sections  we make  some examples concerning the application of Weyl-ordered  basis {\em a l\`a} Bender-Dunne  to basic problems pertinent the integrability of quantum non-autonomous systems whenever the condition (\ref{a5}) on the potential is fulfilled, such as in the cases where the dependence of potential on position is through a power-law term or a series whose time-dependent coefficients scale by a factor at the same power of the coordinate one.
Our investigation will be confined to the former case. In particular, the general formulation of the problems that we will analize  is as follows: 
Let a non-autonomous hamiltonian system (\ref{a2}) be given and let $I\equiv I(q,p,t) $ be an invariant operator for the system. Which is the operator $\Theta\equiv \theta(q,p,t) $ formally conjugated to $I$ via the commutator equation 
\beq
[I,\Theta]=i\hbar\,\, ,
\label{c4}
\eeq
and how can we express $\Theta$ in a Weyl-ordered basis of position- and momentum-type operators? 
 
To answer this query, we will be actuated by the discussions and remarks given in the previous Sections. In particular,   solutions to (\ref{c4}) will be derived in the time-dependent basis.  
 
We  start the discussion of examples by considering the time dependent Hamiltonian 
\beq
H=\frac{p^2}{2m(t)}+\frac{1}{2}m(t)\,\, \omega(t)^2\,\, q^2 \,\, , \eeq
which can be mapped into the new non-autonomous Hamiltonian 
\beq
H^{'}=\frac{1}{2m\rho^2}H_0 \,\, , \qquad H_0=P^2+Q^2 \,\, , 
\label{h0}
\eeq
by means of a linear canonical transformation of the type (\ref{a1}). An invariant operator 
$I(q,p,t)= U H_0 U^{-1}$ can  be devised contextually which  reads 
\beq
I(q,p,t)= 
\rho^2\,p^2-m\,\rho\,\dot{\rho}\,(pq+qp)+(m^2\dot{\rho}^2+\rho^{-2})\,q^2 \,\, 
\,\, .
\eeq
The operator function $\Theta(q,p,t)$ canonically conjugated to $I(q,p,t)$ can be found as a series expansion in the $T_{r,s}$ basis,   
\beq
\Theta (q,p,t)=\sum_{r,s}\alpha_{r,s}(t)\,\, T_{r,s}, 
\label{rec2}
\eeq
The resulting recursion relation for the coefficients $\alpha_{r,s}$ in (\ref{rec2}) is
\beq
\rho^2(s+1)\alpha_{r-1,s+1}-(m^2\dot{\rho}^2+\rho^{-2})(r+1)\alpha_{r+1,s-1}-
(s-r)\,m\,\rho\,\dot{\rho}\,\alpha_{r,s}=\frac{1}{2}\,\delta_{r,0}\delta_{s,0} \,\, \,.
\label{ap1}
\eeq
We look for the \emph{minimal solution} to (\ref{ap1}), where by \emph{minimal solution} we mean  the solution with starting point $\alpha_{-1,1}= 1/(2\rho^2)$ and $\alpha_{r,s}=0$ for $r\leq 2$ and $\forall\,s$. 
In doing so, we first reduce the recursion relation (\ref{ap1}) to a one-index recursion relation with the substitution $s=k=-r$. Redefining the coefficients as
\beq
A_k =\alpha_{r-1,\,s+1}=\alpha_{-k-1,\,k+1} \,\, , 
\label{ap2}
\eeq
 equation (\ref{ap1}) for the $A_k$'s  thus becomes
  \beq
 \rho^2 (k+1)A_k+(m^2\dot{\rho}^2+\rho^{-2})(k-1)A_{k-2}-2\,k\,m\,\rho\,\dot{\rho}\, A_{k-1}=0\,\,,
 \label{ap3}
 \eeq
with initial condition $A_0=1/(2\rho^2)$. 
Equation  (\ref{ap3}) turns into a recursion relation with constant coefficients after the substitution  $A_k=\frac{B_k}{(k+1)}$, i.e.
\beq
 \rho^2\,B_k+(m^2\,\dot{\rho}^2+\rho^{-2})\,B_{k-2}-2\,m\,\rho\,\dot{\rho}\,B_{k-1}=0\,\, ,\qquad B_0=\frac{1}{2\,\rho^2}\,\, .
 \label{ap4}
 \eeq
The idea for solving (\ref{ap4}) is to guess a solution of the form $B_k=x^k$ for some number $x$. To fix the values of $x$ we substitute this expression into (\ref{ap4}) and  identify
 the two roots
\beq
x_\pm=\frac{m(t)\, \rho \,\dot{\rho}\,\pm i}{\rho^2} \,\, .
\label{ap5}
\eeq
The general solution of (\ref{ap4}) is then $B_k=c_+ x_+^k+c_- x_-^k$ with $c_+$ and $c_-$ to be determined by imposing the initial conditions $B_0=\frac{1}{2\rho^2}$ and $B_1=\frac{m(t)\,\dot{\rho}}{\rho^3}$, i.e. 
$$c_{\pm}=\frac{(1 \pm m(t)\,\rho\,\dot{\rho})}{4\, \rho^2} \,\, .$$
The final expression for the operator $\Theta(q,p,t)$ is then
\beq
\Theta(p,q,t)=\frac{ 1}{4\,\rho^2}\sum_{k=0}^\infty A_k \,\, T_{-k,k} 
\label{b1}
\eeq 
with coefficients
\beq
A_k=  \frac{(1+m\,\rho\,\dot{\rho})^{k+1}+(-1)^k(1-m\,\rho\,\dot{\rho})^{k+1}}{(k+1)\,\rho^{2k}}
\label{b2}
\eeq
Instead of solving eq. (\ref{ap1}), we may rely on the results of Sec.\ref{s2}, which suggest {\em i)} to solve the easier recursion relation for the operator 
$\Theta_0(Q,P)=\sum_{m,n} \tilde{\alpha}_{m,n} \,\, \mathcal{T}_{m,n}$
 canonically conjugate to the Hamiltonian $H_0$ in (\ref{h0}) in the  $ \mathcal{T}_{m,n}$ basis
\beq
(n+1)\tilde{\alpha}_{m-1,n+1}-(m+1)\tilde{\alpha}_{m+1,n-1}=\delta_{m,0}\delta_{n,0} \,\, ,
\eeq
 whose solution gives (\cite{B2, BGtime}) 
\beq
\Theta_0 =\sum_{k=0}^{\infty}\frac{(-1)^{k}}{2k+1}\ \cT_{-2k-1,2k+1},
\label{b4}
\eeq
and then {\em ii)}  to express the basis elements $\cT_{-2k-1,2k+1} $ of (\ref{b4})  in terms of the basis elements $T_{m,n}$ in the $(q,p)$-representation obtained from (\ref{E20}) with $n=2k+1$. The series expression for the operator $\Theta$ will then be
\beq
\Theta =\sum_{k=0}^{\infty}\frac{(-1)^{k}}{2k+1}\,\left[ \rho^{-4k-2}T_{-2k-1,2k+1}+
\frac{1}{(m\,\rho\,\dot{\rho})^{2k+1}}\sum_{r=2k+2}^\infty (-1)^r\binom{r-1}{2k} 
\left(\frac{-m\, \dot{\rho}}{\rho}\right)^rT_{-r,r} \right],
\label{b4a}
\eeq
or, by using (\ref{E22}), 
\beq
\Theta =\frac{1}{2}\arctan\left[\frac{X}{\rho^2}\left(1-\frac{m\,\dot{\rho}}{\rho}X\right)^{-1}\right]+\frac{1}{2}\arctan\left[\frac{Y}{\rho^2}\left(1-\frac{m\,\dot{\rho}}{\rho}Y\right)^{-1}\right].
\label{b5}
\eeq
It is easy to check that the coefficients of each $T_{-n,n}$ in (\ref{b1}-\ref{b2}) and (\ref{b4a}) coincide.

\section{Examples. II. Time dependent linear hamiltonian systems }
\label{ss22}
In this Section we shall consider a system ruled by the non-autonomous Hamiltonian
\beq
H=\frac{1}{2}\left[z(t)\, p^{2}+\frac{z(t)}{2\rho(t)^{3}}q\right] \,\, , 
\label{c1}
\eeq
for which the adoption of canonical linear transformation (\ref{a1}) with coefficients (\ref{a7}) 
results into the new Hamiltonian 
 \beq
H'=\frac{z(t)}{2\rho(t)^{2}}H_{0}\,\, ,\qquad {\rm with} \quad H_{0}=P^{2}+Q^{2}+Q \,\,.
\label{c2}
\eeq
So, the hamiltonian formulation of the dynamical problem governed by Equation (\ref{c1}) is 
time-diffeomorphic to an autonomous system whose Hamiltonian is of the form $H_0=P^{2}+Q^{2}+Q$. 
In other words, a genuine harmonic oscillator underlies the case up to a constant shift of position $Q$ 
(and of the energy), $Q\rightarrow\tilde{Q}=Q+1/2$ (and $H_0=P^2+Q^2+Q\rightarrow P^2+\tilde{Q}^2-1/4$, accordingly). 
Solutions to operator dynamical equations having a physical meaning for a system described via (\ref{c1}) can be thus  tailored throughout their corresponding in the harmonic oscillator model. The option of minimal and non-minimal solutions can be argued on similar basis.  
For the sake of completeness and homogeneity with the rest of the paper, and also because 
instructive, in the sequel explicit computation of solution to the 
integrability condition in stationary form for the system (\ref{c1}) will be performed by keeping the operator form (\ref{c2}), 
without reference to the solution for the same problem in the case of purely quadratic Hamiltonian.
Let us start therefore by working out an invariant according to the arguments already expounded.
By operating with the inverse transformation on the Hamiltonian $H_{0}$ in (\ref{c2}), the dynamical invariant 
\beq
I =\rho^{2}(t)\ p^{2}+\beta(t)\ q^{2}-\gamma(t)(pq+qp)\ +\rho^{-1}(t)\ q,
\label{c3}
\eeq
is then obtained for (\ref{c1}),   where 
\beq
\beta(t)=\rho(t)^{-2}+\frac{\dot{\rho}(t)^{2}}{z(t)^{2}} \,\, , \qquad
\gamma(t)=\frac{\dot{\rho}(t)\rho(t)}{z(t)}  \,\, . 
\label{eq: gamma beta}
\eeq 
Seeking a Weyl expansion solution in Bender-Dunne basis $\{T_{m,n}\}$ for the operator $\Theta$ formally conjugated to the operator (\ref{c3}) via Equation (\ref{c4})  therefore leads 
to the difference equation
\beq
\rho(t)^{2}(n+1)\alpha_{m-1,n+1}-\beta(t)\alpha_{m+1,n-1}-\gamma(t)(n-m)\alpha_{m,n}-\frac{1}{2\rho(t)}(m+1)\alpha_{m+1,n}=\frac{1}{2} \delta_{m,0}\delta_{n,0}
\label{c5}
\eeq
for the time-dependent coefficients $\alpha_{m,n}$  connoted in the expansion $\Theta=\sum_{m,n}\alpha_{m,n} \, T_{m,n}$. To the same solution one would arrive by considering 
$H_{0}=P^{2}+Q^{2}+Q$ and solving the quantum integrability condition 
$\left[\Theta_0, H_{0} \right]=i\hbar$ in a Weyl series expansion in the  $\{\cT_{m,n}\}$ basis. 
In particular, assuming $\Theta_0=\sum_{m,n}\tilde{\alpha}_{m,n} \, \cT_{m,n}$  the 
equation $\left[ \Theta_0,\, H_{0} \right]=i\hbar$ is  conveyed onto the difference equation
\beq
2(n+1)\tilde{\alpha}_{m-1,n+1}-2(m+1)\tilde{\alpha}_{m+1,n-1}-(m+1)\tilde{\alpha}_{m+1,n}
=\delta_{m,0}\delta_{n,0} \,\,
\label{c6}
\eeq
for the time-independent expansion coefficients $\tilde{\alpha}_{m,n}$. The minimal solution corresponds to the choice $\tilde{\alpha}_{-1,1}=1/2$ and  $\tilde{\alpha}_{1,-1}=\tilde{\alpha}_{1,0}=0$.
By using the inverse Liouville method expounded in \cite{B3} and summarized in Appendix A,  we guess the general formula of the operator $\Theta$ that is
\beq
\Theta_0=\sum_{k=0}^\infty\sum_{j=0}^k A_{k,j} \,\, \cT_{-2k-1,2k-j+1}
\label{c7}
\eeq 
where the coefficients $A_{k,j}$ must be determined. Equation (\ref{c7}) suggests the following transformation
\beq
A_{k,j}=\tilde{\alpha}_{-2k-1,2k-j+1}=\tilde{\alpha}_{m-1,n+1}
\label{c8}
\eeq
along with the constraint $A_{k,j}=0$ for $k<0$ and $j>k$. 
That means $k=-m/2$ and $j=-m-n,\,(0\leq j\leq k)$.
The new recursion relation then reads
\beq
(2k-j+1)A_{k,j}+(2k-1)A_{k-1,j}+(2k-1)A_{k-1,j-1}=\delta_{k,0}\delta_{j,0} \,\, ,
\label{c9}
\eeq
whose solution is
\beq
A_{k,j}=\frac{(-1)^k}{2k+1}\binom{2k+1}{j} \,\, , 
\label{c10}
\eeq
so that
\beq
\Theta_0=\sum_{k=0}^\infty\sum_{j=0}^k\,\,\frac{(-1)^k}{2k+1}\,\,\binom{2k+1}{j} \,\,
\cT_{-2k-1,2k-j+1}\,\,.
\label{c11}
\eeq
For solving the time-dependent problem, we need to transform the basis elements in (\ref{c11}) according to (\ref{E17}). Remark that in this case  we can also  apply the relation 
\beq
\cT_{-2k-1,2k+1-j}=\frac{i^j (2k+1-j)!}{(2k+1)!} \,\,
\left[ P,\left[P,\dots\left[ P,\cT_{-2k-1,2k+1}\right] \dots\right]\right]_{j-times}\,\, \,\, ,
\label{c12}
\eeq
 and recognize that formally 
 \beq
\cT_{-2k-1,2k+1-j}^{\cC}=\frac{i^j (2k+1-j)!}{(2k+1)!}\,\,
\left[ \rho\, p-m\,\dot{\rho}\, q,\dots\left[ \rho\,p-m\, \dot{\rho}\,q,\cT_{-2k-1,2k+1}^{\cC}\right] \dots\right]_{j-times}\,\, \,\, ,
\label{c13}
\eeq
where
\beq
\cT_{-2k-1,2k+1}^{\cC}=\left(\frac{-m\, \dot{\rho}}{\rho}\right)^{2n+1}T_{-2k-1,2k+1}+\sum_{r=2k+2}^\infty (-1)^r\frac{1}{\rho^{2k+1}}\left(\frac{-m\, \dot{\rho}}{\rho}\right)^r\frac{\Gamma(r+2k+1)}{\Gamma(r+1)\Gamma(2k+1)}T_{-r,r} \,\, .
\eeq

\section{Examples. III. Time-dependent quartic hamiltonian systems}
In this Section we shall detail the analysis of the quantum integrability of a non-autonomous hamiltonian  system associated with the quartic Hamiltonian
\beq
H=\frac{z(t)}{2} \left[ p^2 + \frac{q^4}{2\rho(t)^{6}}\right],
\label{d1}
\eeq
by exploiting the same arguments exposed in previous sections. By application of the canonical coordinate transformation under consideration in this communication, the Hamiltonian (\ref{d1}) can be mapped 
into
\beq
H'=\frac{z(t)}{2\rho(t)^{2}} H_{0} \,\,, \qquad H_0=\frac{P^2+Q^2}{2}+\frac{Q^4}{4} \,\, .
\label{d2}
\eeq
Furthermore, a  dynamical invariant for (\ref{d1}) is obtained which  reads:
\beq
I  =\eta(t)\ p^{2}+\beta(t)\ q^{2}-\gamma(t)(pq+qp)\ +\delta(t)\ q^{4},
\label{d3}
\eeq
where $\eta(t)=\rho(t)^{2}$,  $\delta(t)=\rho(t)^{-4} $,  and 
 $\beta(t)$ and $\gamma(t)$ take the same form as in the linear case
(i.e., $\beta(t)=\rho(t)^{-2}+\frac{\dot{\rho}(t)^{2}}{z(t)^{2}} $ and 
$\gamma(t)=\frac{\dot{\rho}(t)\rho(t)}{z(t)}$; Eq. (\ref{eq: gamma beta})). 
 We are interested in solutions to  the integrability condition $ [\Theta,\,I]= i \hbar $, with $I$ given by equation (\ref{d3}) above. 
Once  a solution is sought in the basis $T_{m,n}$ by means of $\Theta=\sum_{m,n} \alpha_{m,n} \, T_{m,n}$, the recursion relation  
\begin{eqnarray*}
2\eta(t)(n+1)\alpha_{m-1,n+1}-2\beta(t)(m+1)\alpha_{m+1,n-1}-2\gamma(t)(n-m)\alpha_{m,n}\\
-4\delta(t)(m+1)\alpha_{m+1,n-3}+\delta(t)(m+3)(m+2)(m+1)\alpha_{m+3,n-1} & = & \delta(t)(m+3)(m+2)(m+1)\alpha_{m+3,n-1}\end{eqnarray*} 
is obtained for coefficients $\alpha_{m,n}$. 
 With the aim of selecting minimal solution, for $m=n=0$  we choose $\alpha_{1,-1}=\alpha_{1,-3}=\alpha_{3,-1}=0$
and $\alpha_{-1,1}=\frac{1}{2\eta(t)}$. 
With the transformation $M=\frac{1}{6}(2n+m+4K-A)$, $N=\frac{1}{6}(n-m+2K-2A)$, $0\leq A \leq \frac{1}{3}
(n+2)$, $0 \leq K \leq \frac{1}{6}(5n+3m-2)$ we define the new dependent variable 
 \[
C_{N,M,K,A}=\alpha_{-4N+2M-A-1,2N+2M-2K+A+1}=\alpha_{m-1,n+1}\]
along with the constraint that $C_{N,M,K,A}=0$ for $N,M,K,A \leq 0\,,N\leq M\, M\leq K $ and $A \leq M+N-K+1$.
The difference equation for time-dependent coefficients $C_{N,M,K,A}(t)$ arises in the form 
{\small \begin{eqnarray*}
\eta(t)(2M+2N-2K+A+1)C_{N,M,K,A}+\beta(t)(4N-2M+A-1)C_{M-1,N-1,K-1,A}\\
-2\gamma(t)(3N+A-K)C_{N-1,M-1,K-1,A+1}+2\delta(t)(4N-2M+A-1)C_{N-1,M-1,K,A}\\
-\delta(t)(4N-2M+A-1)(4N-2M+A-2)(4N-2M+A-3)C_{N-1,M,K,A} & = & \frac{1}{2}
\delta_{N,0}\delta_{M,0}\delta_{K,0}\delta_{A,0}\end{eqnarray*}
Then 
 \begin{equation}
\Theta =\sum_{N=0}^{\infty}\sum_{M=0}^{N}\sum_{K=0}^{M}\sum_{A=0}^{M+N-K+1}C_{N,M,K,A}(t)\ T_{-4N+2M-A-1,\ 2N+2M-2K+A+1} 
\label{F n a}
\end{equation}
On the other hand, by considering, in the $Q,P$ coordinates, the operator 
$\Theta_0(Q,P) =\sum_{m,n}\tilde{\alpha}_{m,n} \cT_{m,n} $ 
and  the integrability condition $[ H_0,\Theta_0]=i\hbar$
for the stationary quartic-type potential hamiltonian $H_0$ in 
(\ref{d2}),  we get the recurrence constraint
\beq
 (n+1)\tilde{\alpha}_{m-1,n-1}-(m+1)\tilde{\alpha}_{m+1,n-1}+\frac{\hbar^2}{4}(m+1)(m+2)(m+3)\tilde{\alpha}_{m+3,n-1}-(m+1)\tilde{\alpha}_{m+1,n-3} =\delta_{m,0}\delta_{n,0} .
\label{d8}
\eeq	
Equation (\ref{d8}) 
differs by the last term  in the left hand side from the equation one would obtain for the quartic oscillator $H_{BD}=\frac{P^2}{2}+\frac{Q^4}{4}$, the case studied by Bender and Dunne in \cite{B1} (and later also by Galapon in \cite{B3} by exploiting  the Inverse Liouville method).
The structure of the {\em minimal} operator $\Theta_0$ can be written as
\beq
\Theta_0=\sum_{N=0}^{\infty}\sum_{M=0}^{N}\sum_{K=0}^{N-M}A_{MNK} \cT_{-2N-2M-1,\,4N-2M-2K+1} 
\,\, ,
\label{d5bis}
\eeq
as it is also suggested by the inverse Liouville method (see Appendix A). 
By performing the transformation  $m=-2M-2N$ and $n=4N-2M-2K$, the coefficients $A_{M,N,K}$ result connected to the coefficients $\tilde{\alpha}_{m,n}$ through the relation 
\beq
A_{M,N,K}=\alpha_{m-1,n+1}=\tilde{\alpha}_{-2N-2M-1,4N-2M-2K+1}
\label{d9}
\eeq 
 with the constraint  $A_{NMK}=0$ for $N,M,K \leq 0\,,N\leq M\,$ and $ M\leq K $. 
So, } the recursion relation (\ref{d8}) becomes
\begin{align}
\delta_{M,0}\delta_{N,0}\delta_{K,0}&=(4N-2M-2K+1)A_{M,N,K}+(2N+2M-1)A_{M,N-1,K-1}\nonumber \\
&-\frac{\hbar^2}{4}(2N+2M-3)(2N+2M-2)(2N+2M-1)A_{M-1,N-1,K}\nonumber \\
&+(2N+2M-1)A_{M,N-1,k} \,\,.
\label{d10}
\end{align}
A further transformation reduces the partial difference equation -now of the first order in the $N,M,K$'s  variables- to one whose coefficients are linear functions of $N,M,K$. To do this, we define
\beq
B_{NMK} =\frac{2^{-N}\Gamma\left(1/2\right)}{\Gamma\left(N+M+1/2\right)}A_{M,N,K} \,\, , 
\label{d11}
\eeq
along with the constraint that $B_{NMK}=0$ for $N,M,K \leq 0\,,N\leq M\,$ and $ M\leq K $.
This enables us to obtain 
\begin{align}
\delta_{M,0}\delta_{N,0}\delta_{K,0}&=
(4N-2M-2K+1)B_{MNK}+B_{M,N-1,K-1}+B_{M,N-1,K}
 -\hbar^{2}(N+M-1)B_{M-1,N-1,k}\,\, .
\label{d12}
\end{align}
It is simple to derive closed-form expressions for some sets of coefficients, e.g. 
\begin{align*}
 &B_{NN0}  =  \frac{\hbar^{2N}}{1+2N},\quad N\geq0 \,\, ,\\
 &B_{0N0}  =  \frac{(-1)^{N}\Gamma(5/4)}{4^N\Gamma(N+5/4)},\quad N\geq0 \,\, . 
\end{align*}
Finally, a transformation allows to factor the $K$-index dependence out of the coefficients $B_{M,N,K}$. In fact, with the substitution
\beq
B_{M,N,K}=\binom{2N-M+1/2}{K}C_{M,N}
\label{d13}
\eeq
the recursion relation (\ref{d12}) takes the final form
\begin{align}
\delta_{M,0}\delta_{N,0}&=
(4N-2M+1)C_{MN}+C_{M,N-1}
-\hbar^{2}(N+M-1)C_{M-1,N-1},
\label{d14}
\end{align}
that is exactly the recursion relation found in \cite{B2} for the angle variable for the quartic oscillator.  

In the light of the above discussion and of results in \cite{B2}, 
the following statement can be formulated to summarize the traits of the solution to the considered problem:  The {\em minimal} angle-type  operator $\Theta_0$ conjugated via 
$[ H_0,\Theta_0]=i\hbar$ to  the Hamiltonian operator $H_0=P^2+\frac{Q^4}{2}+Q^2$ 
takes the form
\beq
\Theta=\sum_{N=0}^{\infty}\sum_{M=0}^{N}\sum_{K=0}^{N-M}\binom{2N-M+1/2}{K}\,\,C_{M,N}\,\, \cT_{-2N-2M-1,\,4N-2M-2K+1} \,\, 
\label{d15}
\eeq
where the coefficients $C_{m,n}$ can be expressed as
\beq
C_{M,N}=\frac{1}{M!N!}\left( \frac{\partial}{\partial_x}\right)^M \left( \frac{\partial}{\partial_y}\right)^N g(x,y)\vert_{x=y=0},
\label{d16}
\eeq
being $g(x,y)$  the generating function given in \cite{B2}
\beq
g(x,y)=\int_0^1\frac{d\xi}{2\sqrt{\xi}(1-xy\xi)}\exp\left\lbrace \frac{1}{x^2y}\left(\frac{2}{3}-xy+ \frac{(1-xy)^{3/2}(xy\xi-2/3)}{(1-xy\xi)^{3/2}}\right)  \right\rbrace ;
\label{d17}
\eeq
equivalently ($N+M\rightarrow R$), 
\beq
\Theta=\sum_{N=0}^{\infty}\sum_{R=N}^{2N}\sum_{K=0}^{2N-R}
\tilde{C}_{N,R,K}\,\, \cT_{-2R-1,\,6N-4R-2K+1} \,\, \qquad 
\tilde{C}_{N,R,K} =\binom{3N-R+1/2}{K} \,\, C_{R-N,N} \,\, .
\label{d18}
\eeq
Formula (\ref{E17}) can be eventually appealed to write down the formal solution 
$\Theta(q,p,t)$ to the operator equation $[I,\Theta]=i\hbar$ with the invariant $I$ given by (\ref{d3}).

\section{Conclusions}
This communication has been concerned with the problem of solving  operator equations in phase space. In particular, we  entertained  angle-like operators underlying  hamiltonian  systems. For autonomous hamiltonian systems, their defining equation corresponds to the condition that the sought operator is formally conjugated with the Hamiltonian (a condition that, as mentioned in the Introduction, may be seen as an integrability condition for the system, in that at the classical level it ensures the consistency and completeness of the description of evolution and level curves on the time parameter).  
For systems for which a non-autonomous Hamiltonian is given, the condition changes in that the Hamiltonian is no longer the generating function (conserved charge) for time translations. However, while it gets more involved, the defining problem  share analogous technical traits and introduces similar conceptual  
peculiarities.
So far, operator equations in phase space have been approached mainly in regard to 
conservative systems and by considering a Schr\"odinger (i.e. fixed time) 
representation  of the Weyl ordered operator of the Bender-Dunne basis. 
Our aim here has been twofold. From one side, we have been interested in providing and explicit representation of the basis elements in a time-dependent representation. In order to make very manifest the influence of the time-dependence on the realization of the new basis, 
as the class of transformations to superimpose to operators basis in phase space 
we adopted the linear time-dependent canonical ones. We put forward explicit formulas
expressing the relationship between  elements of the Bender-Dunne basis element
resulting in the two canonical pairs representations. Dealing with negative powers of operators involved, these formulas are supplementary to those that can be found in \cite{lohe}.    
Over because of their simplicity, we considered linear time-depending canonical mappings in phase space partially inspired by the line undertook by Mostafazadeh in \cite{mostafazadeh}, as well as by  
the manner in which symplectic time rescaling approaches  are employed for the development of adaptive techniques in molecular dynamics simulations \cite{molec in}-\cite{molec fin}.
By focusing on the simplest of these problems, we next moved to our second scope of solving some concrete  examples of operator equations in phase space that pertain non-autonomous systems. In doing so, we focused on the equations defining angle-like operators conjugated to invariants in some simple non-autonomous cases such as the linear, the quadratic and the quartic type potentials. 
We have performed therefore a comparative study of the treatment of these equations through the 
use of two different basis for the space of solutions, that is the Weyl-ordered basis (\ref{E12}) and the one that results by letting act on it the time-dependent canonical transformation yielding the identification of an invariant. 
Having already elements coming from the discussion of time-dependent quadratic Hamiltonians \cite{B4}, the choice of canonical transformation we made has allowed a prompt comparison and confirmation of formulas derived in section \ref{s2}.
Other than the quadratic ones, time-type operator for linear and quartic non-autonomous one-dimensional hamiltonian  systems have been discussed. 
While selecting explicit solutions by demanding them to concur with \emph{minimal} ones
(the terminology is  borrowed from \cite{B1}), 
a link with the the Inverse Liouville method (see Appendix \ref{appendix liouville}) has been also recalled. For time-dependent quartic Hamiltonian,
the  recurrence relations obtained for the coefficients of the sought minimal angle-like operator put into effect an explicit reappraisal of the operatorial chain Eq. (\ref{coh chain}) that formally expresses solutions obtained by the inverse Liouville solution.   
The minimal solution does not meet the formal structure  of the solutions in the classical limit. This immediately stimulates  the attention on a possible future development of the topic,  because it will be interesting to explicitly desume to what extend the solutions structures suggested by the classical limit are really affected in the semiclassical limit.  To this, one should perform  analysis 
similar to those expounded in previous sections, but incorporating 
the first nontrivial terms in powers of Planck's constant in the approximations of operators $\mathcal{T}_{m,n}$. 
Succeeding in this investigation is expected 
to give new hints on the subject of classical/quantum transitions of dynamical systems.
\bigskip 
\bigskip 
\paragraph*{Acknowledgments.} 
M.G. acknowledges support from the Japan Society for the Promotion of Science 
(Fellowship grant PE14011). 
G.L has been partially supported by 
PRIN 2010 - MIUR "Teorie geometriche e analitiche dei sistemi Hamiltoniani in dimensioni finite e infinite". GL is grateful to the Institute of Industrial Science of the University of Tokyo for the 
hospitality during his visit to perform part of the work.

\section*{Appendix A. Inverse Liouville Method}
\label{appendix liouville}
We summarize here the formal framework presented in \cite{B3} in regard to the treatment 
of the operator equation $ [H, F]=i \hbar$ and its solutions, $H$ denoting a system's Hamiltonian 
operator. The idea upon which the discussion there develops is that, once it is assumed that the inverse of the  operator $\mathcal{L}_{H}=:\left[H,\ \cdot\right]$ exists  (an hypotesis that holds true for analytic potentials), the formal solution for $F$ can be written as $F=i \hbar \mathcal{\hat{L}}_{H}^{-1}$, 
an expression that can be further diversified.
Precisely, if the Hamiltonian is split into its kinetic $K$ and potential $V$ parts, so that $H=K+V$, 
the operator F that formally solves the operator equation under consideration $[H, F]=i \hbar$
(or, in Liouville form,  $\mathcal{L}_{H}\cdot F=i\hbar1$) can be given via the geometric expansion $$F=\sum_{k=0}^{\infty}(-1)^{k} F_{k} \,\,  $$
where $F_0$ denotes the purely kinetical contribution and all the other terms are obtained recursively,
i.e.  
\begin{equation}
F_{0}=\hat{\mathcal{L}_{K}}^{-1}(i\hbar)\,\, , \qquad F_{k}=\left(\hat{\mathcal{L}_{K}}^{-1}\hat{\mathcal{L}}_{V}\right)\, F_{k-1} 
\label{coh chain}\,\, 
\end{equation}
($k\geq 1$).  Being an  operator in the space of pseudo-differentials over $L^2(R)$,
 $\mathcal{L}_{K}$ itself can be defined in terms of its action on Bender-Dunne basis operators. 
Terms in the expansion are accomplished accordingly.

For instance, application of the framework to the quartic Hamiltonian $H=p^{2}+q^{2}+q^{4}$ provides us with the zero-th order term $F_{0}  = -\frac{1}{2}T_{-1,1}$ and the expansion 
\begin{eqnarray}
F & = & -\frac{1}{2}\sum_{k=0}^{\infty}(-1)^{k}\left(\hat{\mathcal{L}_{K}}^{-1}\hat{\mathcal{L}}_{V}\right)^k T_{-1,1} \,\, , \end{eqnarray}
the action of the composite operator $\hat{\mathcal{L}_{K}}^{-1}\hat{\mathcal{L}}_{V}$ on the Bender-Dunne basis being identified by means of the condition
\begin{equation}
\left(\hat{\mathcal{L}_{K}}^{-1}\hat{\mathcal{L}}_{V}\right)T_{-m,n}=-2\left(\frac{m}{n+4}\right)T_{-m-2,n+4}+\frac{1}{2}\frac{m(m+1)(m+2)}{n+2}T_{-m-4,n+2}-\frac{m}{n+2}T_{-m-2,n+2}\label{eq:17}\end{equation}
Repeated applications of equation (\ref{eq:17}) for the specific case $m=n=1$ give rise to 
\begin{equation}
\left(\hat{\mathcal{L}_{K}}^{-1}\hat{\mathcal{L}}_{V}\right)^{n}T_{-1,1}=\sum_{m=0}^{n}
\sum_{k=0}^{m}A_{n,m,k}T_{-4n+2m-1,2n+2m-2k+1} \,\, , \label{eq:18}\end{equation}
for $n\geq0$ and for some constants $A_{n,m,k}$ to be determined. 
It is simple to prove by induction that equation (\ref{eq:18}) holds for all $n\geq0$ and, in the process, 
the recurrence relation that determines uniquely the unknown coefficients $A_{n,m,k}$ with
the boundary condition $A_{0,0,0}=1$ can be rearranged as (\ref{d5bis}).
This proves that the minimal solution coincide with the inverse Liouville solution.

\end{document}